\documentclass[
	aps, 
	prl, 
	twocolumn, 
	superscriptaddress, 
	showpacs,
]{revtex4}

\usepackage{graphicx}
\usepackage{dcolumn}
\usepackage{bm}

\begin{document}


\title{Interplay between carrier and impurity concentrations in annealed Ga$_{1-x}$Mn$_{x}$As intrinsic anomalous Hall Effect}

\author{S.H. Chun}
\affiliation{Department of Physics and Institute of Fundamental Physics, Sejong University, Seoul 143-747, Korea} 
\affiliation{Future Technology Research Division, Korea Institute of Science and Technology, Seoul 130-791, Korea} 
\affiliation{CSCMR, Seoul National University NS 53, Seoul 151-747, Korea} 
\author{Y.S. Kim}
\affiliation{School of Physics and Astronomy, Seoul National University NS 50, Seoul 151-747, Korea} 
\author{H.K. Choi}
\affiliation{CSCMR, Seoul National University NS 53, Seoul 151-747, Korea} 
\affiliation{School of Physics and Astronomy, Seoul National University NS 50, Seoul 151-747, Korea} 
\author{I.T. Jeong}
\affiliation{School of Physics and Astronomy, Seoul National University NS 50, Seoul 151-747, Korea} 
\author{W.O. Lee}
\affiliation{CSCMR, Seoul National University NS 53, Seoul 151-747, Korea} 
\affiliation{School of Physics and Astronomy, Seoul National University NS 50, Seoul 151-747, Korea} 
\author{K.S. Suh}
\affiliation{CSCMR, Seoul National University NS 53, Seoul 151-747, Korea} 
\affiliation{School of Physics and Astronomy, Seoul National University NS 50, Seoul 151-747, Korea} 
\author{\\Y.S. Oh}
\affiliation{CSCMR, Seoul National University NS 53, Seoul 151-747, Korea} 
\affiliation{School of Physics and Astronomy, Seoul National University NS 50, Seoul 151-747, Korea} 
\author{K.H. Kim}
\affiliation{CSCMR, Seoul National University NS 53, Seoul 151-747, Korea} 
\affiliation{School of Physics and Astronomy, Seoul National University NS 50, Seoul 151-747, Korea} 
\author{Z.G. Khim}
\affiliation{School of Physics and Astronomy, Seoul National University NS 50, Seoul 151-747, Korea} 
\author{J.C. Woo}
\affiliation{School of Physics and Astronomy, Seoul National University NS 50, Seoul 151-747, Korea} 
\author{Y.D. Park}
	\email{parkyd@phya.snu.ac.kr}
\affiliation{CSCMR, Seoul National University NS 53, Seoul 151-747, Korea} 
\affiliation{School of Physics and Astronomy, Seoul National University NS 50, Seoul 151-747, Korea} 


\date{\today}

\begin{abstract}
Investigating the scaling behavior of annealed Ga$_{1-x}$Mn$_{x}$As
anomalous Hall coefficients, we note a universal crossover 
regime where the scaling behavior changes from quadratic to linear, 
attributed to the anomalous Hall Effect intrinsic and extrinsic
origins, respectively. Furthermore, measured anomalous Hall conductivities when 
properly scaled by carrier concentration remain constant, equal to
theoretically predicated values, spanning nearly a decade in conductivity as well
as over 100 K in $T_{C}$. Both the qualitative and quantitative agreement 
confirms the validity of new equations of motion including the Berry phase 
contributions as well as tunablility of the intrinsic anomalous Hall Effect.
\end{abstract}

\maketitle

Rudimentary explanation of the Hall Effect, attributed simply to moving 
charge carriers experiencing a Lorentz force, `pressing electricity'\cite{cite1}, 
spectacularly fails to explain even the simplest of the ferromagnetic 
materials. From the original measurements of iron foils by E.H. Hall\cite{cite2} 
to complex correlated oxide systems\cite{cite3} to graphene\cite{cite4}, the familiar 
Hall Effect, from which the nature and amount of carriers can be determined, 
requires an appellation as `ordinary' in a sea of extraordinary effects, 
colorfully termed as `anomalous' to `quantum', and even `extraordinary'. 
Only recently, has the subtle role of the quantum geometry of the Fermi 
surface been recognized as intrinsic origins for much of these 
effects\cite{cite5,cite6}. As any material property determined by transport 
measurements, the Hall Effect reflects contributions from both intrinsic and 
extrinsic mechanisms. Separation of the two and the microscopic origins 
responsible for each has been a source of great contention for decades, 
especially in magnetic materials, and non-reduced dimensioned materials in 
general, with limited experimental observations\cite{cite7,cite8} of a clear intrinsic 
mechanism for the anomalous Hall Effect (AHE). 

In the area of semiconductor spintronics, AHE has had an important role in 
both demonstrating the novelty of carrier mediated ferromagnetic ordering in 
diluted magnetic semiconductors(DMS)\cite{cite9} and indirect characterization
of magnetic properties\cite{cite10,cite11}. Furthermore, a special case of the AHE
with vanishing spin polarization, the intrinsic spin Hall Effect\cite{cite12}, has
recently received much interest as possible sources of spins in a spintronic
device, utilizing the dissipationless nature of the intrinsic transverse spin
current\cite{cite13}. The idea of dissipationless intrinsic Hall current can be
traced to Karplus-Luttinger (KL) formalism\cite{cite14} in which the term `anomalous 
velocity' has been recently reinterpreted as a manifestation of the Berry 
curvature of occupied electronic Bloch states. Following KL arguments, the 
anomalous Hall coefficient ($R_{S}$), related to the strength of the spin 
orbit coupling, scales quadratically with \textit{$\rho$}$_{xx}$, similar behavior to the 
extrinsic side-jump mechanism in which the carrier is asymmetrically 
displaced by impurity scattering, proposed by Berger\cite{cite15}. In between, 
Smit argued that $R_{S}$ must vanish in a periodic lattice and proposed the 
extrinsic skew-scattering mechanism, which predicts the transverse 
resistivity (\textit{$\rho$}$_{xy}$) to scale linearly with \textit{$\rho$}$_{xx}$\cite{cite16}. As much 
experimental evidence showed both linear scaling relationships as well as 
quadratic, in higher resistive samples, AHE for decades have been thought as 
an extrinsic phenomenon related to impurity scattering, and KL ideas of 
intrinsic origins had been discounted. With recent evocation of the Berry 
phase in the momentum space as the intrinsic origins of AHE, in addition to 
predicted scaling relationships, measured \textit{$\sigma$}$_{xy}$ can be directly compared 
to values from well-known material parameters and fundamental quantities to 
discern intrinsic dissipationless spin currents. 

Here, we report a clear and distinct cross-over in the underlying mechanisms 
for the AHE in a series of annealed Ga$_{1-x}$Mn$_{x}$As. In the intrinsic
regime, we observe the transverse conductivity (\textit{$\sigma$}$_{xy}$) when properly
scaled to be independent of longitudinal resistivity (\textit{$\rho$}$_{xx}$), and that
the measured values fit recently proposed theories on the intrinsic origins of AHE evoking
the Berry phase\cite{cite6}. Furthermore, the intrinsic dissipationless spin Hall current, 
impervious to \textit{$\rho$}$_{xx}$, can be manipulated by implicitly and explicitly 
varying the carrier concentration in Ga$_{1-x}$Mn$_{x}$As.

\begin{figure}
\includegraphics[width=85mm]{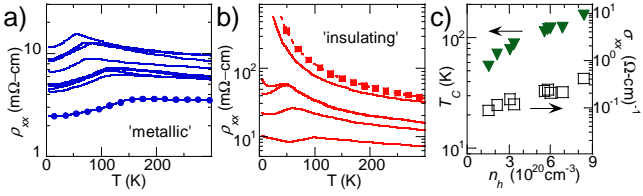}
\caption{(a{\&}b) Series of Ga$_{1-x}$Mn$_{x}$As samples with 
varying $N_{Mn}$ and $N_{Mn}^-$ that exhibit `metallic'-like behavior ((a) circle) and that exhibit 
`insulating'-like behavior ((b) square) are identified by the sign of $\partial \rho_{xx} / \partial T$
far below $T_{C}$. (c) Plot of magnetic ($T_{C}$, triangle) and 
transport (\textit{$\sigma$}$_{xx}$, open square) properties as function of carrier 
concentration ($n_{h}$).}\label{fig1}
\end{figure}

\begin{figure}
\includegraphics[width=85mm]{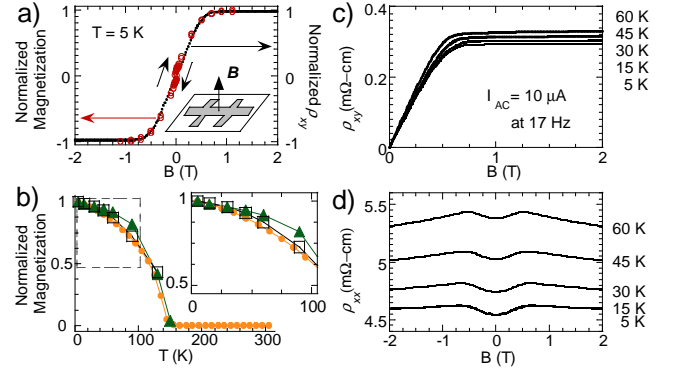}
\caption{(a) AHE (closed circle) reflects SQUID magnetometry 
measurements (open circle) with applied magnetic fields normal to the 
epifilm for sample Ga$_{0.939}$Mn$_{0.061}$As annealed at 250\r{ }C 
(S6.1-250). (b) Normalized magnetization measurements from SQUID (circle); 
from Arrott plots of AHE data (triangle); and from $M \propto \rho_{xy} (\partial \rho_{xy} / \partial B)^{-1}$
or $M \propto \rho_{xy} / R_S$ (open square) (S6.1-250). (c{\&}d) \textit{$\rho$}$_{xx}$
and \textit{$\rho$}$_{xy}$ dependence to applied magnetic field below $T_{C}$
measured simultaneously by ac lock-in techniques (S6.1-250).}\label{fig2}
\end{figure}

Ga$_{1-x}$Mn$_{x}$As, a ferromagnetic semiconductor, is one of the most 
intensively studied materials in the context of semiconductor spintronics, 
and the recipe for its growth is well-known\cite{cite10,cite17}. For our study of the 
AHE in Ga$_{1-x}$Mn$_{x}$As\cite{cite18,cite19}, we explicitly vary the total Mn 
concentration ($N_{Mn}$) between 2.4 and 6.1{\%} during low-temperature 
molecular beam epitaxial growth (LT-MBE). Furthermore, we implicitly vary 
other impurity concentrations by low temperature annealing with temperatures 
ranging from 200-350\r{ }C. After growth, the samples are fashioned into 
electrically isolated 300 $\mu$m $\times$ 1900 $\mu$m Hall bar structures. 
Samples are then annealed in a tube furnace in a flowing dry N$_{2}$ 
environment for one hour with annealing temperature measured by a 
thermocouple near the sample. After annealing, indium contacts are fashioned
and verified as ohmic for transport measurements (Figs. \ref{fig1}{\&}\ref{fig2}).

Similar to reports by others\cite{cite20,cite21}, we observe the magnetic properties, 
in terms of the highest magnetic ordering temperatures ($T_{C}$), and 
transport properties, in terms of lowest resistivities, to be optimized at 
an annealing temperature of $\sim$250\r{ }C with further deterioration of 
such properties at higher annealing temperatures. Both our transport and 
magnetic measurements are consistent with a widely held view that low 
temperature annealing initially removes Mn$_{I}$ ($N_{Mn}^{++}$)\cite{cite22} 
and other electrically active donor impurities as evidenced by increases in 
longitudinal conductivity (\textit{$\sigma$}$_{xx}$), hole carrier concentration ($n_{h}$), 
and T$_{C}$. Annealing above optimal temperatures results in deterioration 
of such properties as Mn$_{Ga}$($N_{Mn}^-$), the source of both spins and 
carriers, decreases\cite{cite21}. Even for the highest annealing temperature 
considered (350\r{ }C), magnetization data along with high resolution x-ray 
diffraction \textit{$\theta$}-2\textit{$\theta$} measurements do not show secondary ferromagnetic phases. However, the 
existence of such phases cannot be fully ruled out\cite{cite19}. In short, after 
LT-MBE and low temperature annealing, a series of Ga$_{1-x}$Mn$_{x}$As 
samples exhibiting both insulator- and metallic-like behaviors with \textit{$\sigma$}$_{xx}$ 
ranging more than two orders of magnitude, $n_{h}$ ranging
from $\sim$0.7-$\sim$8 $\times$ 10$^{20}$cm$^{-3}$, and $T_{C}$ ranging from $\sim$25 K to 
$\sim$160 K was obtained (Fig. \ref{fig1}.c).

\begin{figure}
\includegraphics[width=85mm]{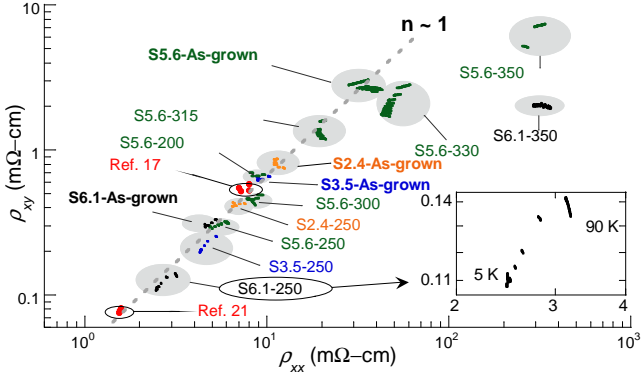}
\caption{For each annealed Ga$_{1-x}$Mn$_{x}$As with differing
$N_{Mn}^-$ and $N_{Mn}^{++}$, variations of \textit{$\rho$}$_{xx}$
and \textit{$\rho$}$_{xy}$ by temperature and applied magnetic 
fields below respective $T_{C}$'s are plotted along with data from two other 
groups (ref. 17 {\&} 21). Each cluster of data points represents an isotherm 
of a particular annealed
Ga$_{1-x}$Mn$_{x}$As (inset for S6.1-250) ($n \approx$ 1 line as a guide).}\label{fig3}
\end{figure}

Plotting all the measured \textit{$\rho$}$_{xx}$ and \textit{$\rho$}$_{xy}$ (Fig. \ref{fig3}), the data indicate an 
existence of an overall scaling relationship akin to the skew scattering 
origins for the AHE even though \textit{$\rho$}$_{xx}$ reflects a wide variance of impurity 
concentrations. The scaling relationship (\textit{$\rho$}$_{xy}$ = c\textit{$\rho$}$_{xx}^{n}$) as 
applied to Ga$_{1-x}$Mn$_{x}$As is problematic, as magnetization is 
inevitably related to \textit{$\rho$}$_{xx}$ by $n_{h}$ as well as weak localization of 
carriers for higher resistive samples\cite{cite23}; thus, the relationship 
$R_{S}$ = c\textit{$\rho$}$_{xx}^{n}$ may be more appropriate. It has been widely reported 
that the magnetization of Ga$_{1-x}$Mn$_{x}$As, even for metallic samples, 
does not easily saturate, thus making a differentiation between the ordinary 
term and the anomalous term difficult. Thus, we estimate $n_{h}$ by fitting 
the temperature dependence of \textit{$\rho$}$_{xy}$\cite{cite24}. With the anisotropy of the 
Ga$_{1-x}$Mn$_{x}$As epilayer in relation to the applied magnetic field and 
with $R_{S}$ much larger than the ordinary Hall coefficient ($R_{o}$), we 
determine $R_{S}$ as $(\partial \rho_{xy} / \partial B)(\mu_o \partial M_Z / \partial B_Z)^{-1}$
from $\rho_{xy} = R_o B + \mu_o R_S M$, an empirical 
relationship which is upheld for both intrinsic and extrinsic origins of 
AHE. For the range of $x$ considered in this study, we expect $\partial M_Z / \partial B_Z$ to
be constant ($\sim$1/\textit{$\mu$}$_{o}$) from anisotropic 
arguments along the lines of Liu \textit{et al.} and Titova \textit{et al.}\cite{cite25}. These
two studies find the perpendicular uniaxial anisotropy fields to be nearly independent of $n_{h}$ and
the cubic anisotropy terms to be negligible due to the large built-in
strain during LT-MBE. Thus, we estimate \textit{cR}$_{S}$
as $\partial \rho_{xy} / \partial B$ (Fig. \ref{fig2}.b).

\begin{figure}
\includegraphics[width=85mm]{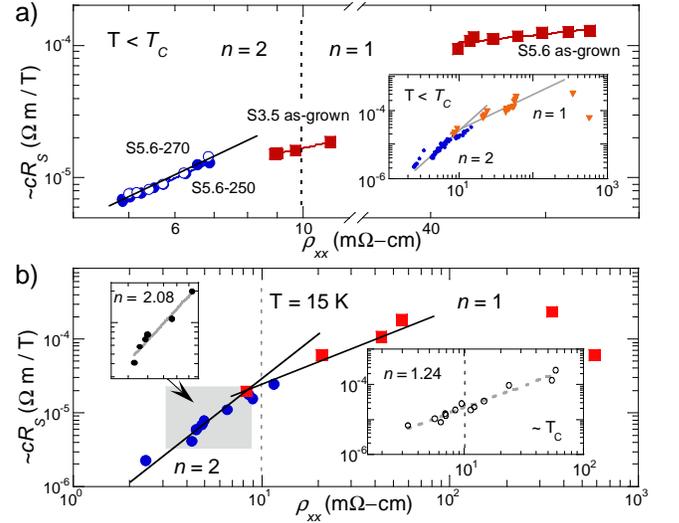}
\caption{(a) \textit{cR}$_{S}$ ($\propto \left. {\partial \rho_{xy} / \partial B} \right.$) dependence
to temperatures below $T_{C}$ (each data point represents 
$R_{S}$ estimated from AHE measurements at T) and \textit{$\rho$}$_{xx}$ are plotted for 
four different samples: two exhibiting quadratic and two linear behaviors. 
\textit{cR}$_{S}$ for all samples for all measurement temperatures below sample's 
$T_{C}$ (inset). (b) For a measurement temperature
of 15 K, \textit{cR}$_{S}$ dependence
to \textit{$\rho$}$_{xx}$ is plotted for all samples with differing 
$N_{Mn}^-$ / $N_{Mn}^{++}$. Lines representing $n$ = 1 and $n$ = 2 are provided as a guide. Quadratic scaling 
behavior indicative of intrinsic AHE is found below 10 m$\Omega$-cm and 
linear above. Plotting similar dependence near T$_{C}$ for each particular 
sample show an overall linear-like scaling relationship (inset).}\label{fig4}
\end{figure}

\begin{figure}
\includegraphics[width=85mm]{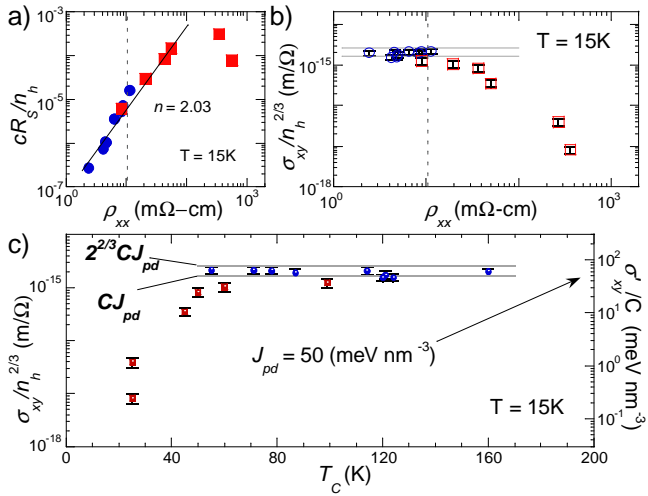}
\caption{(a) Normalized \textit{cR}$_{S}$ (at T = 15 K) to $n_{h}$ show a 
quadratic dependence to \textit{$\rho$}$_{xx}$ except for the two highest resistive samples 
($n$ = 2.03). (b{\&}c) Normalized \textit{$\sigma$}$_{xy}$ (measured at T = 15 K) to 
$n_h^{2/3}$ for `metallic'-like samples fits to the `JNM' lower bound value \textit{CJ}$_{pd}$ for 
wide ranging \textit{$\rho$}$_{xx}$ (b) and T$_{C}$'s (c).}\label{fig5}
\end{figure}

Plotting log(\textit{cR}$_{S}$) vs. log(\textit{$\rho$}$_{xx}$) (Fig. \ref{fig4}.a) shows a clear demarcation 
where a fit to $n$ in the scaling relationship ($R_{S}$ = c\textit{$\rho$}$_{xx}^{n}$) 
changes from a value of two to one, for \textit{$\rho$}$_{xx} >$ 10 m$\Omega$-cm as 
similarly alluded by Ruzmetov \textit{et al.}\cite{cite24}, especially at lower temperatures 
(Fig. \ref{fig4}.b). Here, we vary both temperature and magnetic solute 
concentrations ($N_{Mn}$) to vary \textit{$\rho$}$_{xx}$. Furthermore, for temperatures 
near $T_{C}$ (Fig. \ref{fig4}.b inset), the scaling relationship seems to converge to 
a value of $n$ equal to one, a result analogous to skew scattering origins of 
AHE in paramagnetic matrix with embedded magnetic impurities\cite{cite26}, or 
possibly due to phonon-assisted hopping of holes between localized states in 
the impurity band\cite{cite23,cite27}. For some of the highest annealing temperatures, 
we note $n >$ 3, suggestive of inhomogeneous systems\cite{cite19}. Whether the 
dissipationless anomalous Hall current is intrinsic for all 
Ga$_{1-x}$Mn$_{x}$As samples is not clear by examining the scaling 
relationships alone. Lee \textit{et al.} report of the dissipationless intrinsic AHE in 
the ferromagnetic spinel CuCr$_{2}$Se$_{4-x}$Br$_{x}$ maintains that a 
quadratic relationship between $R_{S}$/$n_{h}$ and c\textit{$\rho$}$_{xx}$ should be 
expected\cite{cite8}. Interestingly, a similar plot (Fig. \ref{fig5}.a) indicates quadratic 
relationship between $R_{S}$/$n_{h}$ and \textit{$\rho$}$_{xx}$ for all samples 
considered except for two samples annealed at the highest temperature, 350\r{ }C. 

The unexpected quadratic relationship between $R_{S}$/$n_{h}$ and \textit{$\rho$}$_{xx}$
for Ga$_{1-x}$Mn$_{x}$As questioned us what should be the proper 
normalization method for Ga$_{1-x}$Mn$_{x}$As. The answer comes from the 
intrinsic AHE theory developed by Jungwirth, Niu and MacDonald (JNM)\cite{cite6}. 
In brief, they use a semi-classical transport theory including the effect of 
Berry phase:
\begin{equation}
\label{eq1}
\dot{\vec{x}} = \frac{1}{\hbar} \frac{\partial E_n (\vec{k})}{\partial \vec{k}} + \dot{\vec{k}} \times \vec{B}_n (\vec{k})
\end{equation}
\begin{equation}
\label{eq2}
\hbar \dot{\vec{k}} = -e(\vec{E} + \dot{\vec{x}} \times \vec{B} (\vec{x}))
\end{equation}
The resemblance between (Eq. 1) and (Eq. 2) is why the Berry
curvature $\vec{B}_n (\vec{k})$ is called ``the magnetic field in momentum space''. JNM 
calculated $\sigma_{xy}$ \textit{exactly} with the assumptions of infinite spin-orbit 
coupling strength and a mean-field Hamiltonian with the effective field $h$ is 
given by $N_{Mn} SJ_{pd}$, where $N_{Mn}$ is the density of Mn ions with 
spin $S = 5/2$:
\begin{equation}
\label{eq3}
CN_{Mn} J_{pd} n_p^{-1/3} < \sigma_{xy} < 2^{2/3} CN_{Mn} J_{pd} n_p^{-1/3} 
\end{equation}
where $C = \frac{5}{2} \frac{e^2}{2\pi \hbar} \frac{(3\pi^2)^{-1/3}}{2\pi \hbar^2} m_{hh}$. The
upper and lower bounds of $\sigma_{xy}$ correspond to 
$m_{lh} \approx m_{hh}$ and $m_{lh} << m_{hh}$, respectively, with $\sigma_{xy}$
of Ga$_{1-x}$Mn$_{x}$As being closer to the lower bound ($m_{lh} / m_{hh} = 0.16$). In
JNM calculations, there exists a tacit condition for a 
`clean-limit' where all of $N_{Mn}$ participate in the magnetic ordering as well as 
later inclusions of finite spin-orbit coupling; warping of band structures; 
and strains and defects in applying to experimental data\cite{cite28}.

Again, in our scheme to vary \textit{$\rho$}$_{xx}$, we explicitly varied $N_{Mn}$ during 
LT-MBE growth and implicitly varied $N_{Mn}^{++} $ and $N_{Mn}^-$ by low 
temperature annealing. In course of determining simple scaling relationships 
of $R_{S}$ to \textit{$\rho$}$_{xx}$, we observe much of the transport properties and 
magnetic properties to have a distinct dependence on $n_h$ (Fig. \ref{fig1}.c), as 
expected from a carrier mediated DMS as Ga$_{1-x}$Mn$_{x}$As\cite{cite10}. Moriya 
and Munekata found that $N_{Mn}^-$ becomes saturated despite the steady 
increase of $N_{Mn}$, and consistent scattering coefficients when $N_{Mn}^-$
was used in the room-temperature AHE analysis\cite{cite29}. Magnetization study 
also supports this notion: the seemingly deficient magnetization is 
recovered if only the ionized Mn atoms are counted\cite{cite30}.

Therefore, it is concluded that $N_{Mn}^-$ or $n_{h}$ rather than $N_{Mn}$ 
characterizes Ga$_{1-x}$Mn$_{x}$As epilayers. The monotonic dependences of 
$T_{C}$ and $\sigma_{xy}$ on $n_{h}$ inferred further support the argument. 
Then, we find that (Eq. 3) is enough to explain the behavior of metallic 
Ga$_{1-x}$Mn$_{x}$As samples once we replace $N_{Mn}$ with $n_h$. Equation 
3 then simplifies to: $\sigma_{xy} = CJ_{pd} n_h^{2/3}$ if we take the 
lower bound. Thus, we normalize $\sigma_{xy}$ by $n_h^{2/3}$ instead of $n_{h}$ and 
the results are shown (Figs. \ref{fig5}.b{\&}c). Clearly the two classes (metallic 
and insulating) of Ga$_{1-x}$Mn$_{x}$As express differing behaviors with the 
same crossover as the change in scaling behavior of $R_{S}$. While the 
insulating samples show drastic changes as $T_{C}$ varies, the metallic 
samples show a good scaling behavior despite the large change in $T_{C}$ from 
50 K to 160 K. The most striking observation is the excellent quantitative 
agreement with $CJ_{pd}$, when we use the widely accepted value of $J_{pd}$ = 50 meV nm$^{-3}$. 

To summarize, our data clearly shows a universal crossover between intrinsic 
and extrinsic in the scaling behavior of the anomalous Hall coefficient and 
that anomalous Hall conductivities of metallic Ga$_{1-x}$Mn$_{x}$As follows 
the theoretical prediction qualitatively and quantitatively. And, we find 
the intrinsic nature of AHE, akin to dissipationless spin Hall currents, in 
Ga$_{1-x}$Mn$_{x}$As to be robust.

\begin{acknowledgments}
We will like to thank T.F. Ambrose, Y.B. Kim, S.J. Pearton, and M.B. Salamon 
for their critical readings of the manuscript. We also like to thank the 
MSL at KBSI for allowing access to MPMS. This work 
is partly supported by KOSEF and Samsung Electronics Endowment through 
CSCMR. S.H.C. is partly supported by KIST Vision 21 program. K.H.K and 
Y.D.P. is partly supported by the City of Seoul R{\&}BD Program. 
\end{acknowledgments}

\end{document}